\documentclass[conference]{IEEEtran}
\IEEEoverridecommandlockouts
\usepackage{bbm}
\usepackage{times}
\usepackage[cmex10]{amsmath}
\usepackage{amssymb}
\usepackage{epsfig,verbatim}
\usepackage{algorithm,algpseudocode}
\usepackage{hhline}
\usepackage{algpseudocode}
\usepackage{booktabs}
\usepackage{multirow}
\usepackage{graphicx, subfigure}
\DeclareGraphicsExtensions{.pdf,.png,.jpg}
\usepackage{cite}
\usepackage{mathtools}
\usepackage{epsfig,latexsym,cite}
\usepackage{setspace}
\usepackage{multicol}
\usepackage{color}
\usepackage{cuted}
\usepackage{stfloats}

\def\BigRoman{\uppercase\expandafter{\romannumeral\number\count 255 }}
\def\Romannumeral{\afterassignment\BigRoman\count255=}

\long\def\comment#1{}

\newfont{\bbb}{msbm10 scaled 700}

\newfont{\bb}{msbm10 scaled 1100}
\newcommand{\CC}{\mbox{\bb C}}

\newcommand{\av}{{\bf a}}
\newcommand{\bv}{{\bf b}}

\newcommand{\hv}{{\bf h}}

\newcommand{\mv}{{\bf m}}

\newcommand{\rv}{{\bf r}}
\newcommand{\sv}{{\bf s}}

\newcommand{\xv}{{\bf x}}
\newcommand{\yv}{{\bf y}}
\newcommand{\zv}{{\bf z}}

\newcommand{\Am}{{\bf A}}

\newcommand{\Em}{{\bf E}}

\newcommand{\Hm}{{\bf H}}

\newcommand{\Mm}{{\bf M}}

\newcommand{\Cc}{{\cal C}}

\newcommand{\Pc}{{\cal P}}

\newcommand{\alphav}{\hbox{\boldmath$\alpha$}}

\newcommand{\SNR}{{\sf SNR}}

\renewcommand{\Re}{{\rm Re}}
\renewcommand{\Im}{{\rm Im}}
\newcommand{\eqdef}{\stackrel{\Delta}{=}}

\newcommand{\transp}{{\sf T}}

\newtheorem{definition}{Definition}

\newtheorem{remark}{Remark}

\newcommand{\argmax}{\operatornamewithlimits{argmax}}

\title{Low-Complexity Symbol-Level Precoding for MU-MISO Downlink Systems with QAM Signals}

\author{Sungyeal Park,~\IEEEmembership{Student,~IEEE,} Yunseong Cho,~\IEEEmembership{Student,~IEEE,}
        and~Songnam Hong,~\IEEEmembership{Member,~IEEE}
\thanks{S. Park is with Artificial Intelligence Convergence Network, Ajou University, Suwon, Korea (e-mail: awdrg1541@ajou.ac.kr)}
\thanks{Y. Cho is with the Department of Electrical and Computer Engineering, The University of Texas Austin, TX, USA (e-mail: yscho@utexas.edu)}
\thanks{S. Hong is with the Department of Electronic Engineering, Hanyang University, Seoul,  Korea (e-mail:snhong@hanyang.ac.kr)}        
}

\begin{document}
\maketitle


\begin{abstract}
This study proposes the construction of a transmit signal for large-scale antenna systems with cost-effective 1-bit digital-to-analog converters in the downlink. Under quadrature-amplitude-modulation constellations, it is still an open problem to overcome a severe error floor problem caused by its nature property. To this end, we first present a feasibility condition which guarantees that each user's noiseless signal is placed in the desired decision region. For robustness to additive noise, we formulate an optimization problem, we then transform the feasibility conditions to cascaded matrix form. We propose a low-complexity algorithm to generate a 1-bit transmit signal based on the proposed optimization problem formulated as a well-defined mixed-integer-linear-programming.
Numerical results validate the superiority of the proposed method in terms of detection performance and computational complexity.
\end{abstract}


\section{Introduction}\label{sec:intro}

Recently, massive multiple-input single-output (MISO) systems have been actively investigated as a core technology in fifth-generation (5G) and future wireless communication systems due to its significant gain in spectral efficiency \cite{marzetta2010noncooperative}. 
One of the key challenges is dealing with a high hardware cost caused by a large number of radio frequency (RF) chains which consist of nearly linear power amplifiers (PA) and digital-to-analog converters (DACs) for each antenna element. In massive MISO systems, the total power consumption at BS is increased by the number of the RF chains. 
Moreover, in downlink systems, the power consumption from the PAs and DACs accounts for the majority of the total power consumption at BS. 
Therefore, the use of power-efficient low-resolution DACs has gathered momentum as a promising low-power solution in a variety of application spaces \cite{spencer2004introduction,larsson2014massive,Choi2021Quantized}. 
In traditional downlink systems, zero-forcing (ZF) and regularized ZF (RZF) achieve almost optimal performance effectively \cite{peel2005vector}.
These linear precoding schemes with low complexity are widely used in wireless communication with nearly linear PAs and high-resolution DACs (e.g., 12 bits). Unfortunately, power consumption grows exponentially with the number of quantization bits. 
For this reason, massive MISO must be built with low-cost DACs. 

Non-linear precoding methods are based on various design criteria such as minimum mean square error (MMSE), constructive Interference (CI), maximum safety margin (MSM). For phase-shift-keying (PSK) and quadrature-amplitude-modulation (QAM) constellations, the methods for MMSE criterion are introduced in \cite{castaneda20171,jacobsson2016nonlinear,sohrabi2018one,jacobsson2017quantized,chen2019mmse,wang2018finite}. 
In \cite{castaneda20171}, C1PO, C2PO method is proposed as a low-complexity algorithm variant from using bi-convex relaxation. Unfortunately, these methods do not provide good performance with  QAM constellations. 
In \cite{jacobsson2016nonlinear}, non-linear 1-bit precoding methods are enabled by semi-definite relaxation and $\ell_\infty$-norm relaxation. However, these methods do not provide an elegant complexity-performance trade-off. 
In \cite{chen2019mmse}, The MMSE-based one-bit precoding, MMSE-ERP is developed by a combination of the alternating minimization method using a projected gradient method, and equilibrium constraint. 
The performance of MMSE-ERP is significant. Also, \cite{wang2018finite} provides the IDE algorithm and IDE2 which is a complexity-efficient algorithm of IDE that exploits an alternating direction method of multipliers (ADMM) framework. Both IDE and IDE2 achieve excellent error-rate performance. 

The CI design criterion is similar to our one. \cite{li2020interference,landau2017branch,li20201bit,li2018massive} propose symbol-level precoding methods that utilize CI design criterion. In \cite{landau2017branch}, a precoding method based on branch-and-bound (B\&B) is proposed in the massive MIMO systems with PSK constellations. Symbol scaling (SS) is the low-complexity algorithm that achieves good performance with PSK. 
In \cite{li20201bit,li2020interference}, a partial branch and bound (P-BB) and an ordered partial sequential update (OPSU), based on the optimization problems defined with both equality constraints and inequality constraints achieve near-optimal performance and significant performance, respectively.
In \cite{jedda2018quantized}, MSM design criterion exploiting the CI and MSM algorithm and analysis of the algorithm are provided for constant envelope precoding with PSK and QAM.
These methods do not provide an elegant complexity-performance trade-off since in QAM constellations, the decision regions are bounded. Thus, The major subject of this paper is to investigate a precoding method with a near-optimal performance and low complexity under QAM constellations.

In this paper, we design a novel direction to construct a 1-bit transmit signal vector for a downlink MU-MISO system with 1-bit DACs. A first key contribution is the so-called {\em feasibility condition} which guarantees that the noiseless observation of each user belongs to a desired decision region. If a transmit signal vector satisfy the feasibility condition, each users can detect a desired signal at high signal-to-noise ratio (SNR).
To combine the robustness to an additive noise into the feasibility condition, we transform our problem as a mixed integer linear programming (MILP), which can be optimally solved via B\&B. Furthermore, we present a low-complexity method to solve the MILP via a novel greedy algorithm, which yield the near optimal performance. Via numerical results, we show that the proposed method perform state-of-the-art performances. Moreover, the potential of the presented direction and methods is demonstrated by a run-time comparison of the 1-bit precoding methods.

This paper is organized as follows. We represent useful notations and definitions, and describe a system model in Section~\ref{sec:pre}. In Section \ref{sec:structure}, we propose an design criterion using the feasibility condition to construct a transmit signal vector for downlink MU-MISO systems with 1-bit DACs. Moreover, in Section \ref{sec:low complexity methods}, the low complexity method are proposed. Section \ref{sec:simulation} demonstrates numerical results. Conclusions are provided in  \ref{sec:conclusion}. 
\section{Preliminaries}\label{sec:pre}

In this section, we provide useful notations used throughout the paper, and then describe the system model.

\subsection{Notation}
The uppercase and lowercase bold letters represent matrices and column vectors, respectively. For any vector $\bf x$, $x_i$ represents the $i$-th component of $\bf x$. The symbol $(\cdot)^{\transp}$ denotes the transpose of a matrix or a vector. Let $\left[a:b\right]\eqdef\{a,a+1,\ldots,b\}$ for any integer $a$ and $b$ with $a<b$. The notation of $\text{card}(\mathcal{U})$ denotes the number of elements of a finite set  $\mathcal{U}$. A rank of a matrix \Am \ is represented as rank(\Am). 
$\Re(\av)$ and $\Im(\av)$ represent the real and complex parts of a complex vector $\av \in \CC$, respectively.
 For any $x\in\mathbb{C}$, we let
 \begin{equation}\label{eq:1}
     g(x)=[\Re(x), \Im(x)]^{\transp},                          
 \end{equation} 
 and $g^{-1}$ denotes the inverse mapping of $g$. Also, $g$ and $g^{-1}$ are the component-wise operations, i.e., $g([x_1,x_2]^{\transp})=[\Re(x_1),\Im(x_1),\Re(x_2),\Im(x_2)]^{\transp}$. For a complex-value $x$, its real-valued matrix expansion $\phi(x)$ is defined as
 \begin{equation}\label{eq:2}
     \phi(x)=\left[{\begin{array}{cc}
     \Re(x)&  -\Im(x)\\
     \Im(x)&  \Re(x)
     \end{array}}\right].
 \end{equation}
 As an extension to a vector, the operation of $\phi$ is applied in an element-wise manner as
 \begin{equation}
     \phi([x_1,x_2]^{\transp})=[\phi(x_1)^{\transp},\phi(x_2)^{\transp}]^{\transp}.
 \end{equation}
 $\otimes$ indicates Kronecker product operator, and $\bar{\bf{1}}_n$ denotes the length-$n$ all-one vector.


\subsection{System Model}
We consider a downlink of MU-MISO system. The BS with $N_t$ transmits antennas serves $K$ single-antenna users with infinite-resolution ADC, where $N_t \gg K$. $\Cc$ denotes the set of constellation points of $4^n$-QAM with $n\geq 2$. For a standard input-output relation, the received signal vector 
$\yv\in\CC^{K}$ at the $K$ users is given as
 \begin{equation}\label{eq:3}
     \yv = \sqrt{\rho}\Hm\xv+\zv,
 \end{equation}
 where $\xv=[x_1,\ldots,x_{N_t}]^{\transp}$ represents a transmit vector at the BS and  $\Hm\in\mathbb{C}^{K\times N_t}$ denotes the frequency-flat Rayleigh fading channel whose each entry follows a complex Gaussian distribution with zero mean and unit variance. The additive Gaussian noise vector $\zv\in\mathbb{C}^{K\times 1}$ models i.i.d. circularly-symmetric complex Gaussian noise with zero mean and unit variance per each entry, i.e., $z_i \sim \mathcal{CN}(0,\sigma^2=1)$.
 $\rho$ denotes the per-antenna power constraint and the SNR is defined as $\SNR=\rho/\sigma^2$. Throughout the paper, we assume full channel state information (CSI) at the BS.
 
 Given a message vector $\sv\in\mathcal{C}^K$, we propose that BS construct a transmit vector $\xv$ such that each user $k$ can recover the desired message $s_k$ successfully. 
 To this end, we aim to construct a symbol-level precoding function $\Pc$ as
 \begin{equation}\label{eq:4}
     \xv=\mathcal{P}(\Hm,\sv),
 \end{equation} that produces a transmit vector $\xv$ based on $\Hm$ and $\sv$.
According to the one-bit constraint, each component $x_i$ is restricted as 
 \begin{equation}\label{eq:5}
     \Re(x_i)\  \rm{and}\ \Im(\mathit{x_i})\in \{-1,1\}.
 \end{equation} Due to a severe non-linearity from the restriction, conventional methods developed using the linearity cannot guarantee an attractive performance. Our goal is a precoding function $\mathcal{P}(\Hm,\sv)$ with a suitable for the considered non-linearity and manageable complexity.



\section{The Proposed Transmit-Signal Vectors}\label{sec:structure}

In this section, we present an optimization problem that constructs a transmit-vector $\xv$ under $4^n$-QAM.
This problem can be rewritten as a manageable MILP. 
For the ease of exploration, an equivalent real-valued expression is used as
\begin{equation}\label{eq:9}
         \tilde{\yv} = \sqrt{\rho}\tilde{\Hm}\tilde{\xv}+\tilde{\zv},
\end{equation}
where $\tilde{\xv}=g(\xv)$, $\tilde{\xv}=g(\xv)$, $\tilde{\zv}=g(\zv)$, and $\tilde{\Hm}=\phi(\Hm)\in\mathbb{R}^{2K\times 2N_t}$ denotes the real-value matrix of $\Hm$.

First, we provide the useful definitions which are used throughout the paper.

\vspace{0.1cm}
\begin{definition}\label{def1}{\em (Decision region)} For any constellation point $s\in\mathcal{C}$, the decision region of $s$ is defined as
\begin{equation}\label{eq:6}
    \mathcal{R}(s) \triangleq \left\{y\in\mathbb{C}:|y-s| \le \min_{c\in\mathcal{C}:c\ne s}|y-c|\right\}.
\end{equation} A received signal $y$ is deteacted as $s$ if $y$ is in $\mathcal{R}(s)$. Also, the real-valued decision region is given as
\begin{equation}
    \tilde{\mathcal{R}}(s)=g\left(\mathcal{R}(s)\right).
\end{equation}
\end{definition} 

\vspace{0.1cm}
\begin{definition}\label{def2}{\em (Base region)} A base region $\tilde{\mathcal{B}}_i\subseteq \mathbb{R}^2, \forall i\in\left[0:3\right]$, is defined as
\begin{equation}\label{eq:7}
    \tilde{\mathcal{B}}_{i} \triangleq \{\alpha_{i}^1\mv_{i}^1+\alpha_{i}^2\mv_{i}^2: \alpha_{i}^1,\alpha_{i}^2>0\},
\end{equation}
where $\mv_i^\ell$ denotes a basis vector with
\begin{equation}\label{eq:8}
    \mv_{i}^{\ell} = \begin{cases} g\left(\sqrt{2}\cos(\frac{\pi}{4}(1+2i))\right) & \mbox{if }\ell=1 \\ g\left(j\sqrt{2}\sin(\frac{\pi}{4}(1+2i))\right) & \mbox{if }\ell=2.  \end{cases}
\end{equation}
\end{definition} 
\vspace{0.1cm}

\vspace{0.1cm}

 \begin{figure}[!t]
    \begin{center}
    \includegraphics[width=0.35\textwidth]{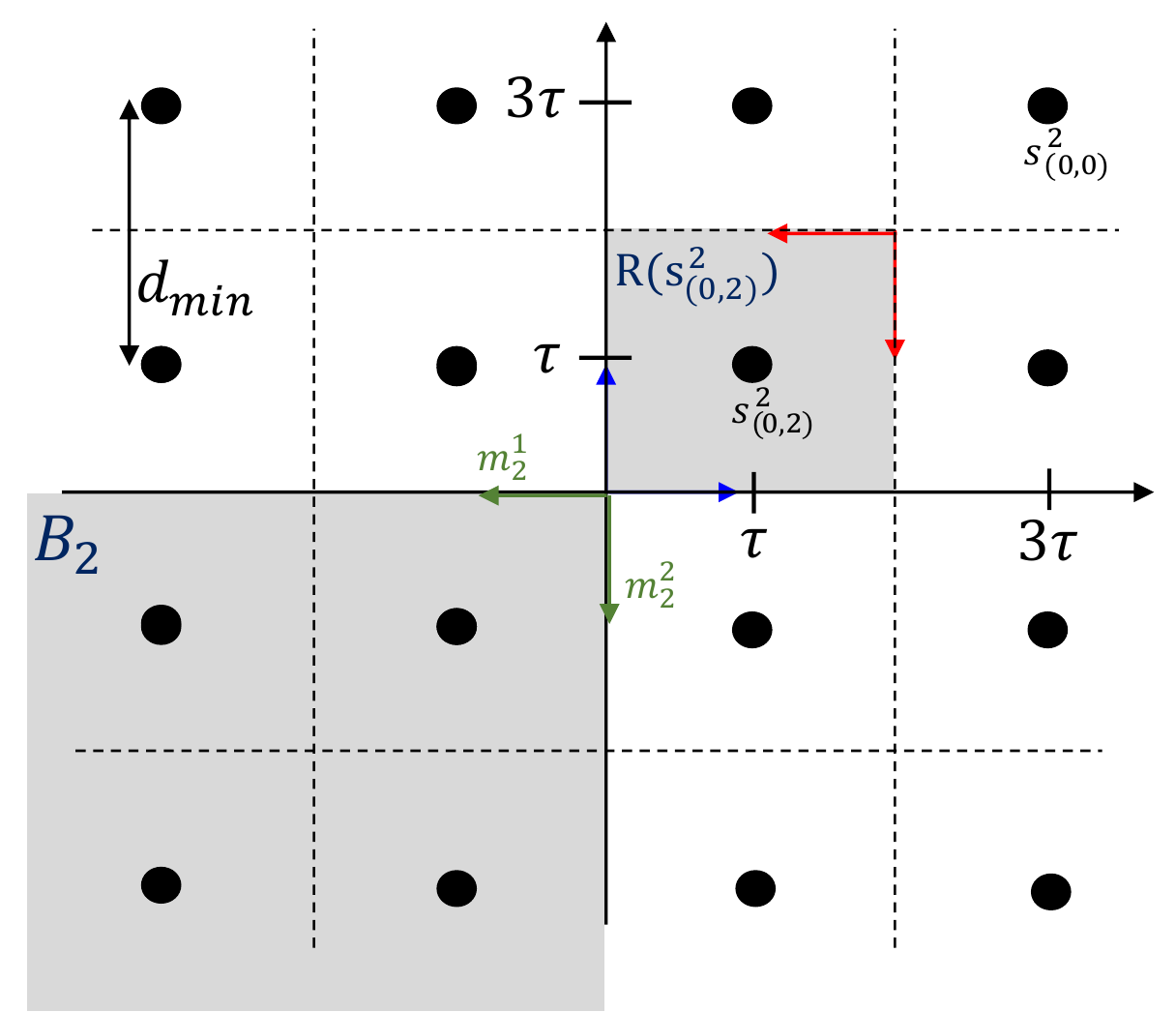}
    \end{center}
    \caption{Description of the decision regions for $4^2$-QAM with adaptive $\tau$.}
    \label{fig:2}
    \vspace{-1.0em}

\end{figure}
We then represent the decision region in Definition~\ref{def1} as an intersection of the $n$ base regions in Definition~\ref{def2} with proper offsets. 
First of all, we need to decide a decision-size $\tau=\frac{d_{\rm min}}{2}$, where $d_{\rm min}$ represents the minimum Euclidean distance of the given constellation points. 
In PSK, $\tau$ is always infinite regardless of a channel, whereas in $4^n$-QAM, it should be optimized. 
Specifically, if a noiseless received signal belongs to the desired decision regions at all the $K$ users, $\tau$ should be as large as possible to guarantee reliable performance. 



From now on, we explain how to construct a transmit-signal vector $\xv$ for a given decision-size $\tau$. Throughout the paper, it is assumed that all $K$ users' decision-size $\tau$ are identical for the practicability of an optimization. Given $4^n$-QAM, each symbol is indexed by a length-$n$ quaternary vector $(i_1,...,i_n)$ with $i_j \in [0:3]$. The following constellation $\mathcal{C}$ and real-valued form $\tilde{\mathcal{C}}$ are respectively represented as
\begin{align}
\mathcal{C}=\left\{s_{(0,\ldots,0)}^{n},s_{(0,\ldots,1)}^{n},\ldots,s_{(3,\ldots,3)}^{n}\right\},\\
\tilde{\mathcal{C}} = \left\{g(s_{(0,\ldots,0)}^{n}),g(s_{(0,\ldots,1)}^{n}),\ldots, g(s_{(3,\ldots, 3)}^{n})\right\}.
\end{align} Each constellation point can be represented as a normalized constellation point, (i.e., a linear combination of the $n$ basis symbols $c_i$'s) with $\tau$ such as 
 \begin{equation}\label{eq:13}
     s_{(i_1,\ldots, i_n)}^{n} \triangleq \tau {s'}_{(i_1,\ldots, i_n)}^{n} = \tau \sum_{l=1}^n 2^{n-l} c_{i_l},
 \end{equation} where $c_i$'s,  the basis symbols are given as
\begin{equation}
    c_i \triangleq \sqrt{2}\left\{{\cos\left({\frac{\pi}{4}} (1+2i)\right)+j\sin\left({\frac{\pi}{4}} (1+2i)\right)}\right\},
\end{equation} 
for $i\in[0:3]$. 
We aims a transmit vector $\xv$ to ensure that a noiseless received signal at the $k$-th user (i.e., $r_k=\hv_k\xv$, where $\hv_k$ is k-th row of $\Hm$) should be placed in the corresponding decision regions for all users $k\in[1:K]$. This essential condition implies that  $\xv$ should satisfy the following condition:
\begin{align}\label{eq:region_constraint}
     g(r_k) &\in \tilde{\mathcal{R}}\left( s^{n}_{(\mu_{k,1},\ldots, \mu_{k,n})}\right),
\end{align} for $k \in [1:K]$.

\vspace{0.1cm}
\noindent{\bf  Feasibility condition:} 
To reform the condition \eqref{eq:region_constraint} as $n$ linear equations, we first represent the decision region in \eqref{eq:region_constraint} as the intersections of the $n$ shifted base regions in Definition~\ref{def2}:
\begin{align}\label{eq:18}
  \tilde{\mathcal{R}}\left( s^{n}_{(i_1, \ldots, i_n)}\right)\triangleq \tilde{\mathcal{B}}_{i_1}\bigcap_{l=2}^n \left\{\tilde{\mathcal{B}}_{i_l}+2^{n-(l-1)}g\left(s_{(i_1,\ldots, i_{l-1})}^{l-1}\right)\right\},
\end{align}where the shifted base region is defined as
\begin{equation}\label{eq:19}
    \tilde{\mathcal{B}}_{i}+c \triangleq \{\alpha_{i}^1\mv_{i}^1+\alpha_{i}^2\mv_{i}^2+c: \alpha_{i}^{1},\alpha_{i}^{2}>0\},
\end{equation}with a bias $c$.  Then, the condition in (\ref{eq:region_constraint}) is established when $g(r_k)$ expressed by the following $n$ linear equations with some positive coefficients, i.e.,
\begin{align}\label{eq:20}
    g(r_k) &= \alpha_{k,1}^1\mv_{\mu_{k,1}}^1+\alpha_{k,1}^2\mv_{\mu_{k,1}}^2+2^{n}g(0) \\
            & = \alpha_{k,2}^1\mv_{\mu_{k,2}}^1+\alpha_{k,2}^2\mv_{\mu_{k,2}}^2+2^{n-1}g(s_{(\mu_{k,1})}^{1}) \nonumber  \\
            &\;\; \vdots  \nonumber \\
            & = \alpha_{k,n}^1\mv_{\mu_{k,n}}^1+\alpha_{k,n}^2\mv_{\mu_{k,n}}^2+2^1g(s_{(\mu_{k,1},\ldots ,\mu_{k,n-1})}^{n-1}), \nonumber
\end{align} for some $\alpha_{k,1}^1,\alpha_{k,1}^2, \ldots,\alpha_{k,n}^1,\alpha_{k,n}^2 \ge 0$.
The condition in (\ref{eq:20}) is called a {\em feasibility} condition. 
Satisfying the condition ensures that all $K$ users can detect the desired messages in the high SNR regime, i.e., $r_k\in \mathcal{R}\left( s^{n}_{(\mu_{k,1},\ldots,\mu_{k,n})}\right)$ for $k\in [1:K]$.


We now represent the feasibility condition in a matrix form. Define the $n$ copies of the channel vector $\hv_k$ as
\begin{equation}
    \Hm^k\eqdef \bar{\bf{1}}_n \otimes \hv_k=[\underbrace{\hv_k^\transp,\ldots, \hv_k^\transp}_{n}]^\transp. \label{eq:26} 
\end{equation}
The corresponding real-valued channel is represented as 
\begin{equation}
    \tilde{\Hm}^k=\phi(\Hm^k).
\end{equation} Accordingly, the $n$-extended received vector of $k$-th user is denoted as
\begin{align}\label{eq:27}
    \rv^k&\triangleq g(\Hm^k\xv)= \tilde{\Hm}^k\tilde{\xv} =\bar{\bf{1}}_n \otimes g(r_k).
\end{align} Then, The right-hand side of \eqref{eq:20}, i.e., linear constraints, is represented in a matrix form.  Using Definition 2, we let:
\begin{align}
    \Mm_i \triangleq [\mv_{i}^1 \ \mv_{i}^2]&= \begin{bmatrix} \Re(c_{i}) & 0 \\ 0 & \Im(c_{i})   \end{bmatrix},
   \label{eq:25}    
\end{align} which is a orthogonal and symmetric matrix.
For whole message of $k$-th user, $\mu_k$, we now pack each message terms from $n$ linear equations \eqref{eq:20} to a more manageable format.
The basis matrix $\Mm^{\mu_k}$, coefficient vector $\alphav^k$ are respectively represented as
\begin{align}
    \Mm^{\mu_k} \triangleq \mbox{diag}(\Mm_{\mu_{k,1}},\ldots,\Mm_{\mu_{k,n}}), \label{eq:28}\\
\alphav^k \triangleq[\alpha_{k,1}^1,\alpha_{k,1}^2,\ldots,\alpha_{k,n}^1,\alpha_{k,n}^2]^\transp.
\end{align}
Also, the normalized bias vector $\bv$ of k-th user's all biases with $\tau$ is given as
 \begin{align}\label{eq:29}
    \bv^{\mu_k} \triangleq g\left([2^n\cdot0, 2^{n-1}\cdot {s'}_{(\mu_{k,1})}^{1}, \ldots ,2^1\cdot {s'}_{(\mu_{k,1},\ldots,\mu_{k,n-1})}^{n-1}]^\transp\right)\nonumber\\
    =\frac{1}{\tau}g\left([2^n\cdot0, 2^{n-1}\cdot s_{(\mu_{k,1})}^{1}, \ldots ,2^1\cdot s_{(\mu_{k,1},\ldots,\mu_{k,n-1})}^{n-1}]^\transp\right).
 \end{align}
Using \eqref{eq:28}-\eqref{eq:29}, the $k$-th user's feasibility conditions \eqref{eq:20} is given as matrix equation,
\begin{equation}\label{eq:31}
   \rv^k=\Mm^{\mu_k}\boldsymbol{\alpha}^k+\tau\bv^{\mu_k}.
\end{equation}
The cascaded matrix form of feasibility conditions for all $K$ users is constructed as
\begin{equation}\label{eq:32}
    \Bar{\rv}=\Bar{\Hm}\tilde{\xv}=\Bar{\Mm}\bar{\boldsymbol{\alpha}}+\tau\Bar{\bv},
\end{equation}
where 
\begin{gather}
        \Bar{\Mm} \triangleq \mbox{diag}(\Mm^{\mu_1},\ldots,\Mm^{\mu_K}), \\
        \Bar{\Hm} \triangleq [(\tilde{\Hm}^{1})^\transp,\ldots,(\tilde{\Hm}^{K})^\transp]^\transp, \\
 \Bar{\rv} \triangleq [(\rv^{1})^\transp,\ldots,(\rv^{K})^\transp]^\transp,\\
 \Bar{\bv} \triangleq [(\bv^{\mu_1})^\transp,\ldots,(\bv^{\mu_K})^\transp]^\transp \\
        \bar{\boldsymbol{\alpha}}\triangleq [(\boldsymbol{\alpha}^1)^\transp,\ldots,(\boldsymbol{\alpha}^K)^\transp]^\transp.
\end{gather} 
Leveraging the fact that $\bar{\Mm}^{-1}=\bar{\Mm}$ from (\ref{eq:25}), the feasibility condition in (\ref{eq:32}) is rewritten as
\begin{equation}\label{eq:37}
    \bar{\boldsymbol{\alpha}}=\underbrace{\bar{\Mm}\bar{\Hm}}_{\triangleq{\boldsymbol{\Lambda}}}\tilde{\xv}-\tau\underbrace{\bar{\Mm}\bar{\bv}}_{\triangleq{\boldsymbol{\Lambda}_b}}.
\end{equation} 
\noindent{\bf Robustness:} Unfortunately, a feasible transmit vector cannot guarantee robustness to the additive Gaussian noise despite providing attractive performance in higher SNR.
Thus, we formulate an optimization problem aiming to move away the noiseless signal from the boundaries of the decision areas as 
\begin{align}\label{eq:39}
 &\mathcal{P}_1:&& \max_{\tilde{\xv},\tau} \min\{\alpha_{k,j}^i: i=1,2,\ j\in[1:n],\ k\in[1:K]\} \\
  & \text{s.t.} &&\bar{\boldsymbol{\alpha}}=\boldsymbol{\Lambda}\tilde{\xv}-\tau\boldsymbol{\Lambda}_{b},\nonumber\\
  &&& \alpha_{k,j}^1,\alpha_{k,j}^2 > 0,\ j\in[1:n],\ k\in[1:K],\nonumber\\
   &&& \tilde{\xv}\in\{-1,1\}^{2N_t} \nonumber.
\end{align}
\begin{figure}[!t]
    \centering
    \includegraphics[width=0.410\textwidth]{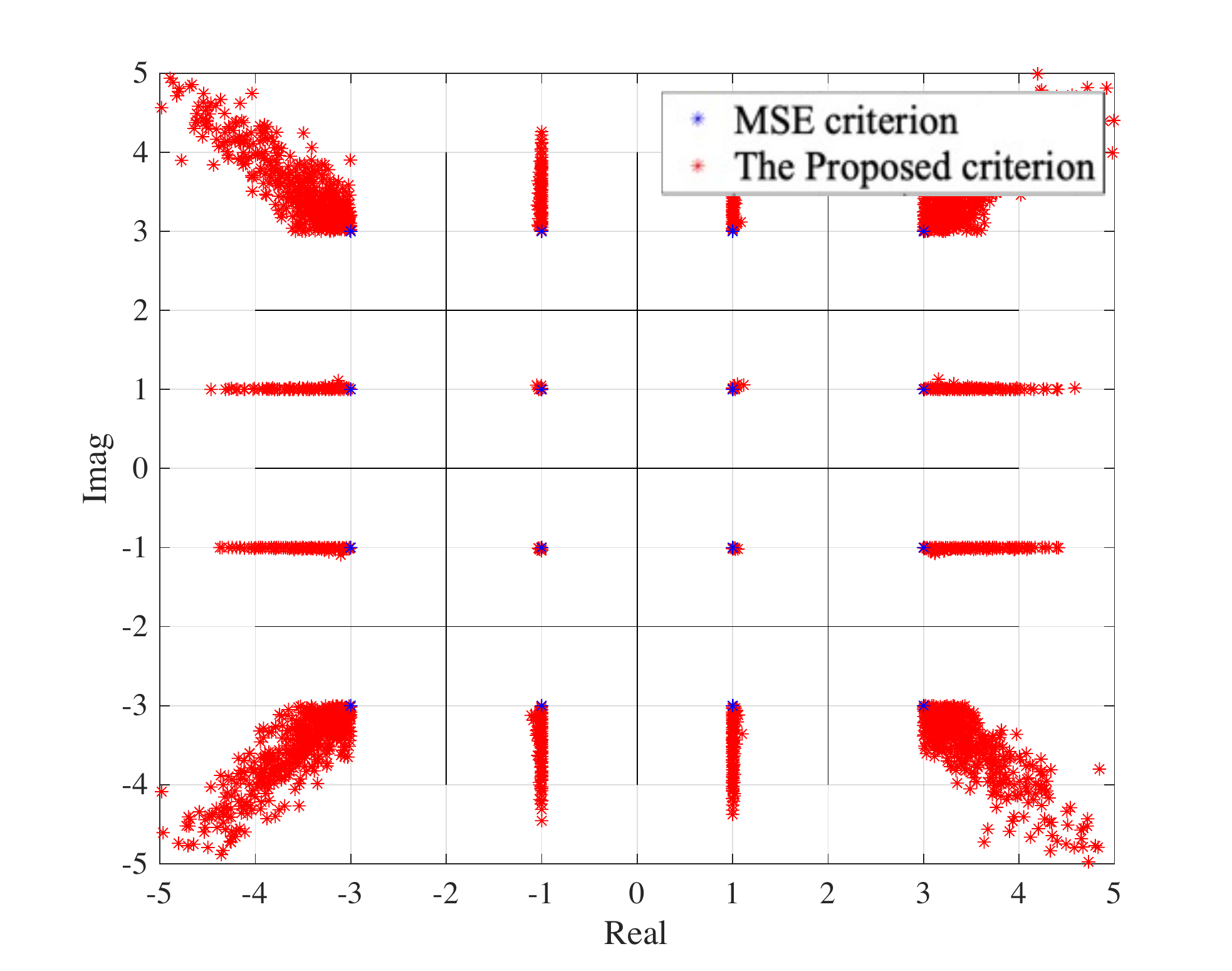}
    \caption{The normalized noiseless received signals of $4^2$-QAM.}
    \label{fig:4}
    \vspace{-1.0em}

\end{figure} 
To solve the problem $\mathcal{P}_1$ efficiently, we express $\mathcal{P}_1$ as MILP: 
\begin{align}\label{eq:43}
 &\mathcal{P}_2:&& \argmax_{\tilde{\xv},t}\ \  t \nonumber\\
  & \text{s.t.} &&\boldsymbol{\Lambda}_i\tilde{\xv}-\tau \boldsymbol{\Lambda}_{b,i} \ge t,\ i\in[1:2nK], \nonumber\\ 
  &&& t>0, \nonumber\\
   &&& \tilde{\xv}\in\{-1,1\}^{2N_t},
\end{align} where $\boldsymbol{\Lambda}_i$ and $\boldsymbol{\Lambda}_{b,i}$ represent the $i$-th row of $\boldsymbol{\Lambda}$ and $\boldsymbol{\Lambda}_b$, respectively.
We remark the fact that the object function $t$ is the maximized lower bound of the coefficients $\bar{\alphav}$ from $\mathcal{P}_2$. The coefficients directly indicate how far away it is from a detection boundary. Based on this facts, we set $t$ to proper $\tau$, i.e., $\tau \eqdef t$.
Accordingly, the MILP problem to the decision-size $\tau$ and transmit vector $\xv$ simultaneously is defined as
\begin{align}\label{MILP4}
 &\mathcal{P}_3:&& \argmax_{\tilde{\xv},t}\ \  t \nonumber\\
  & \text{s.t.} &&\frac{1}{1+\boldsymbol{\Lambda}_{b,i}}\boldsymbol{\Lambda}_i\tilde{\xv} \ge t,\ i\in[1:2nK], \nonumber\\ 
  &&& t>0, \nonumber\\
   &&& \tilde{\xv}\in\{-1,1\}^{2N_t}.
\end{align} 
Although the widely used B\&B method for the MILP can solve the proposed MILP problem in $\mathcal{P}_3$ \cite{landau2017branch}, this method is not appropriate due to its infeasible complexity in realistic implementation \cite{landau2017branch}. 


\begin{remark} Fig.~\ref{fig:4} clearly shows the proposed approach, where $10^4$ normalized noiseless signals, i.e.,  $\Hm\xv$, are plotted with $N_t=8$, $K=2$, and $4^2$-QAM.
The blue points describe the noiseless received signals from unquantized transmit vectors using ZF precoding without 1-bit constraint in \cite{peel2005vector}. 
In contrast, the red points depict the noiseless received signals from the proposed 1-bit transmit vectors, i.e., the solutions of $\mathcal{P}_3$. Fig.~\ref{fig:4} demonstrates that the red points can provide more robustness than the blue points even with the low-resolution data converters.
\end{remark}

\section{Low-Complexity Precoding Method}\label{sec:low complexity methods}
In this section, we propose efficient algorithm to solve MILP problems in $\mathcal{P}_3$. 
First of all, the integer constraint in $\mathcal{P}_3$ is relaxed as the bounded interval not to loose convexity:
\begin{align}\label{eq:46}
 &\mathcal{P}_4:&& \argmax_{\tilde{\xv},t}\ \  t \nonumber\\
  & \text{s.t.} &&\frac{1}{1+\boldsymbol{\Lambda}_{b,i}}\boldsymbol{\Lambda}_i\tilde{\xv} \ge t,\ i\in[1:2nK], \nonumber\\ 
  &&& t>0, \nonumber\\
  &&& -1\le\tilde{x}_j\le 1,\ j\in[1:2N_t].
\end{align}

The relaxed LP problem, $\mathcal{P}_4$ can be solved efficiently via simplex method \cite{luenberger1984linear}. Here, $\tilde{\xv}_{\rm LP}$ denotes the solution of $\mathcal{P}_4$. We refine $\tilde{\xv}_{\rm LP}$ obtained by $\mathcal{P}_4$ to satisfy the 1-bit constraints via a full greedy algorithm, which is summarized in Algorithm 1.
Furthermore, we note that finding $\tilde{\xv}_{\rm LP}$ via simplex method explores an extreme point of constraint set of $\mathcal{P}_4$. Due to the fact that extreme points are basic feasible solutions, most entries of $\tilde{\xv_{\rm LP}}$ already satisfy 1-bit constraint. 
In addition, pivoting of the simplex method depends on the rank of standard LP's constraint matrix. From \cite{park2021construction}, the rank$(\boldsymbol{\Lambda})$ is equal to the rank of LP constraints of the proposed method, which is $2K$. Therefore, complexity of LP almost depends on the number of users $K$. To verify the fact, we demonstrate the run-time simulation in Fig.~\ref{fig:runtime}.

\begin{algorithm}[t]\label{al:1}
\caption{Greedy Algorithm}
\textbf{Input:}  $\tilde{\xv}_{\rm LP}\in\mathbb{R}^{2N_t\times1}$, $\boldsymbol{\Lambda}\in\mathbb{R}^{2nK\times2N_t}$, $\boldsymbol{\Lambda}_b\in\mathbb{R}^{2nK\times1}$ and $\tau\in\mathbb{R}^{+}$.

\textbf{Initialization:} $\tilde{\xv}=\tilde{\xv}_{\rm LP}$ (obtained by $\mathcal{P}_4$).
\begin{algorithmic}
\For{$i=1:2N_t$}
\For{$j\in \{-1,1\}$}
\State $\tilde{x}_i=j$ and $\bar{\boldsymbol{\alpha}}^{(j)}=\boldsymbol{\Lambda}\tilde{\xv}-\tau \boldsymbol{\Lambda}_b$
\EndFor
\State  Update $\tilde{x}_i\leftarrow \argmax_{j\in\{-1,1\}}\{\min(\bar{\boldsymbol{\alpha}}^{(j)})\}$
\EndFor
\State \textbf{Output:} $\tilde{\xv}\in\mathcal{R}^{2N_t\times1}$
\end{algorithmic}
\end{algorithm}
%



\section{Numerical Results}\label{sec:simulation}

In this section, we validate the superiority of the proposed method over existing methods in terms of symbol-error-rate (SER) and computational complexity.

\begin{figure}[!t]
    \centering
    \includegraphics[width=0.41\textwidth]{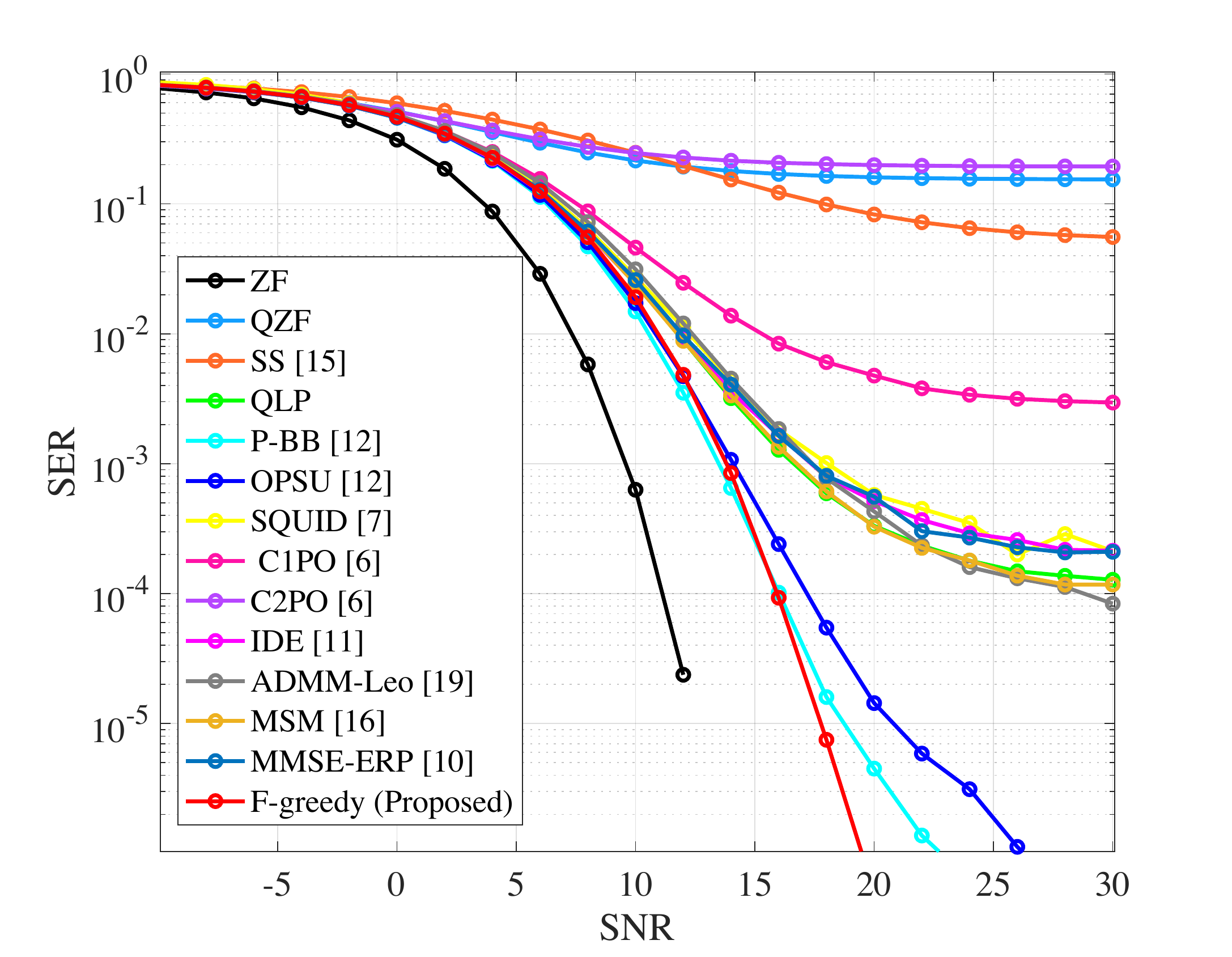}
    \caption{Performance comparisons of precoding methods for the downlink MU-MISO systems with 1-Bit DACs, where $N_t$=64, $K$=8, and $4^2$-QAM with adaptive $\tau$.}
    \label{fig:7}
\end{figure}

\begin{figure}[!t]
    \centering
    \includegraphics[width=0.41\textwidth]{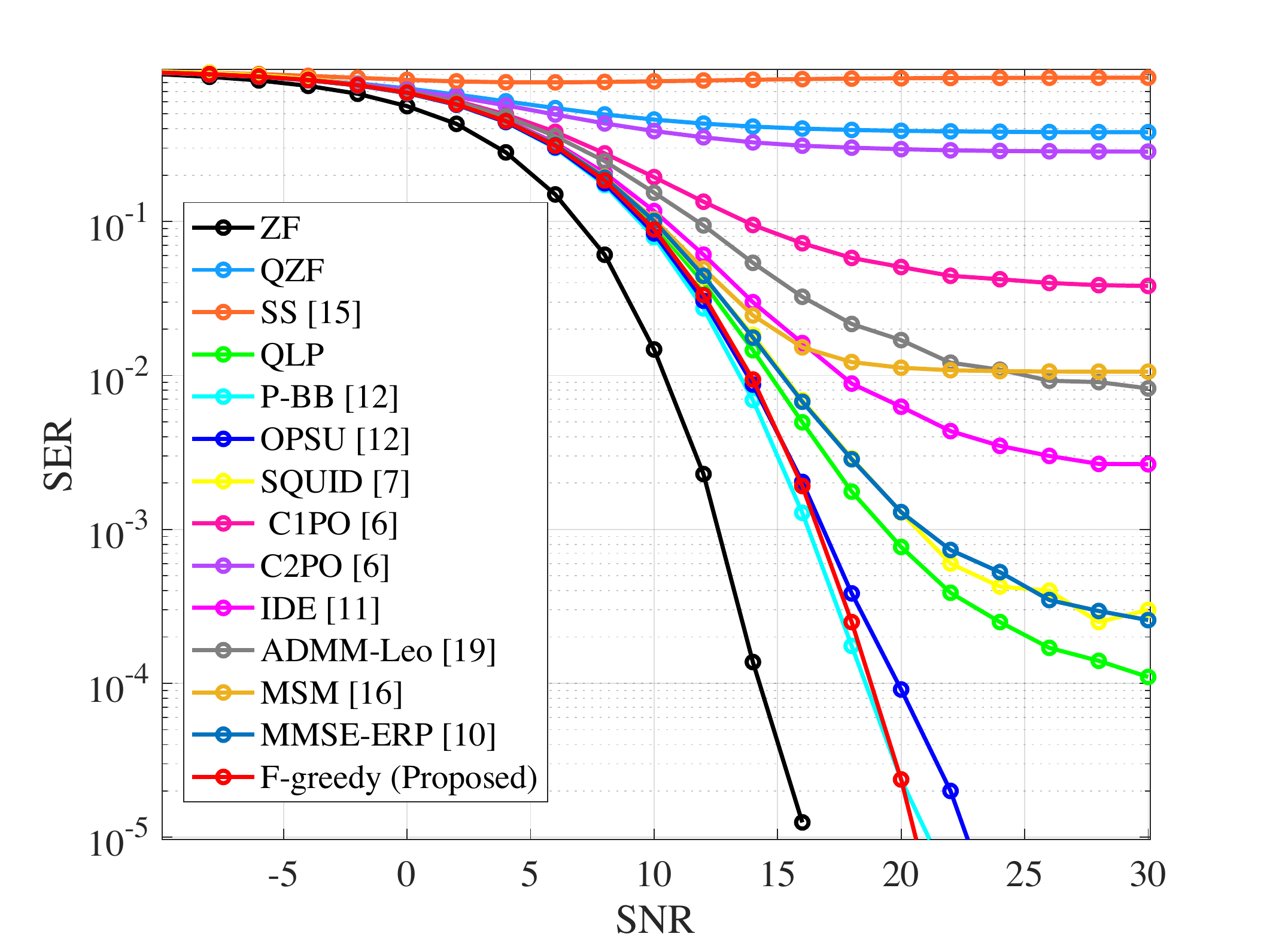}
    \caption{Performance comparisons of precoding methods for the downlink MU-MISO systems with 1-Bit DACs, where $N_t$=128, $K$=8, and $4^3$-QAM with adaptive $\tau$.}
    \label{fig:8}
    \vspace{-1.5em}

\end{figure}

In Figs.~\ref{fig:7} and \ref{fig:8}, {\bf ZF} is the conventional ZF method with infinite-resolution DACs by the lower-bound of the 1-bit precoding methods. Quantized zero forcing ({\bf QZF}) and Quantized LP ({\bf QLP}) are the direct 1-bit quantization of ZF and 1-bit quantization of the solution from $\mathcal{P}_4$, respectively. Also, the efficient and excellent 1-bit precoding methods, such as {\bf SS} \cite{li2018massive}, {\bf P-BB, OPSU} \cite{li2020interference}, {\bf SQUID} \cite{jacobsson2016nonlinear}, {\bf C1PO, C2PO} \cite{castaneda20171}, {\bf IDE} \cite{wang2018finite}, {\bf ADMM-Leo} \cite{chu2019efficient}, {\bf MSM method} \cite{jedda2018quantized} and {\bf MMSE-ERP} \cite{chen2019mmse}  are compared with the proposed method, full greedy(namely, {\bf F-greedy}) based LP from Algorithm 1.
Respectively, all benchmarks follow parameter settings in \cite{li2018massive, li2020interference, jacobsson2016nonlinear, castaneda20171, wang2018finite, chu2019efficient, jedda2018quantized, chen2019mmse} throughout the simulations.

Fig.~\ref{fig:7} presents the performance comparisons of the MU-MISO case with $N_t=64$, $K=8$, $4^2$-QAM, and adaptive $\tau$. An optimal performance is obtained from the ZF methods with infinite-resolution data converters, which is interpreted as the lower-bound of the 1-bit methods. Unfortunately, due to unfeasible complexity, the performance of MILP cannot be evaluated.
At high SNR, except for LP-based methods, most 1-bit precoding methods including {\bf QLP} (i.e., solving $\mathcal{P}_4$) suffer from a severe error-floor.
Therefore, to maintain the feasibility and robustness with 1-bit constraint, we add the proposed algorithm \ref{al:1} (namely, F-greedy).
The proposed method achieves the near-optimal performance, which implies that $\tau$ from the $\mathcal{P}_4$ is close to optimal. 
We note that the P-BB and OPSU methods search fewer candidates than our algorithm, thereby having a minor performance loss.
In detail, our optimization problem can express all candidates in the decision region as an intersection of $n$ base regions with inequality constraints only, however the P-BB and OPSU methods include equality constraints as well, which diminish the search space. 

In Fig.~\ref{fig:8}, we observe the same aspect of the systems, where $N_t=128$, $K=8$, and $4^3$-QAM with adaptive $\tau$.
Unlike most methods including P-BB and OPSU that find $\tau$ alternatively, the $\tau$ is fixed at once by the proposed method. The rationality of the $\tau$ from $\mathcal{P}_4$ is observed in Figs.~\ref{fig:7} and \ref{fig:8}.
\begin{figure}
    \centering
    \includegraphics[width=0.41\textwidth]{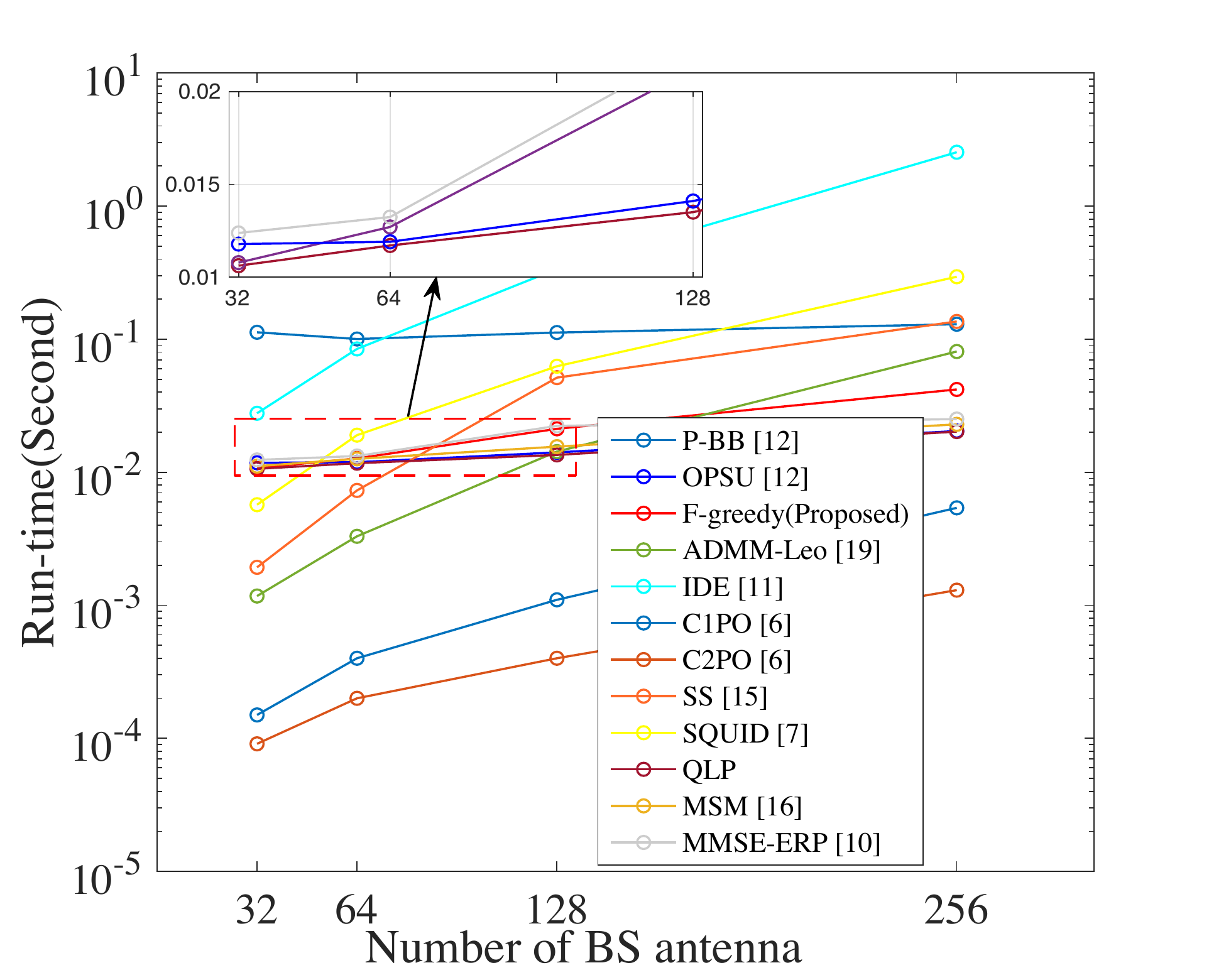}
    \caption{Run-time versus the number of BS antennas for precoding methods , where $K$=8, and $4^2$-QAM with adaptive $\tau$.}
    \label{fig:runtime}
    \vspace{-1.5em}

\end{figure}



Fig.~\ref{fig:runtime} shows the run-time comparison of the methods with $10^4$ simulations. In Fig.~\ref{fig:runtime}, the novelty of our algorithm is demonstrated when having large-scale antennas arrays. In detail, run-time of the proposed method is about 10 times less than P-BB, but the SER performance turn out to be the same as P-BB with near-optimal performance. 
In addition, since the computational complexity of the simplex method that solves $\mathcal{P}_4$ mainly hinges on the number of users, the run-time of the proposed method growing with number of antennas is caused by greedy algorithm \ref{al:1}, which track all entries of the transmit vector. Although this problem, the run-time of our method is still quite small. the complexity problem is solved in \cite{park2021construction}.

\begin{figure}[!t]
     \centering
     \includegraphics[width=0.41\textwidth]{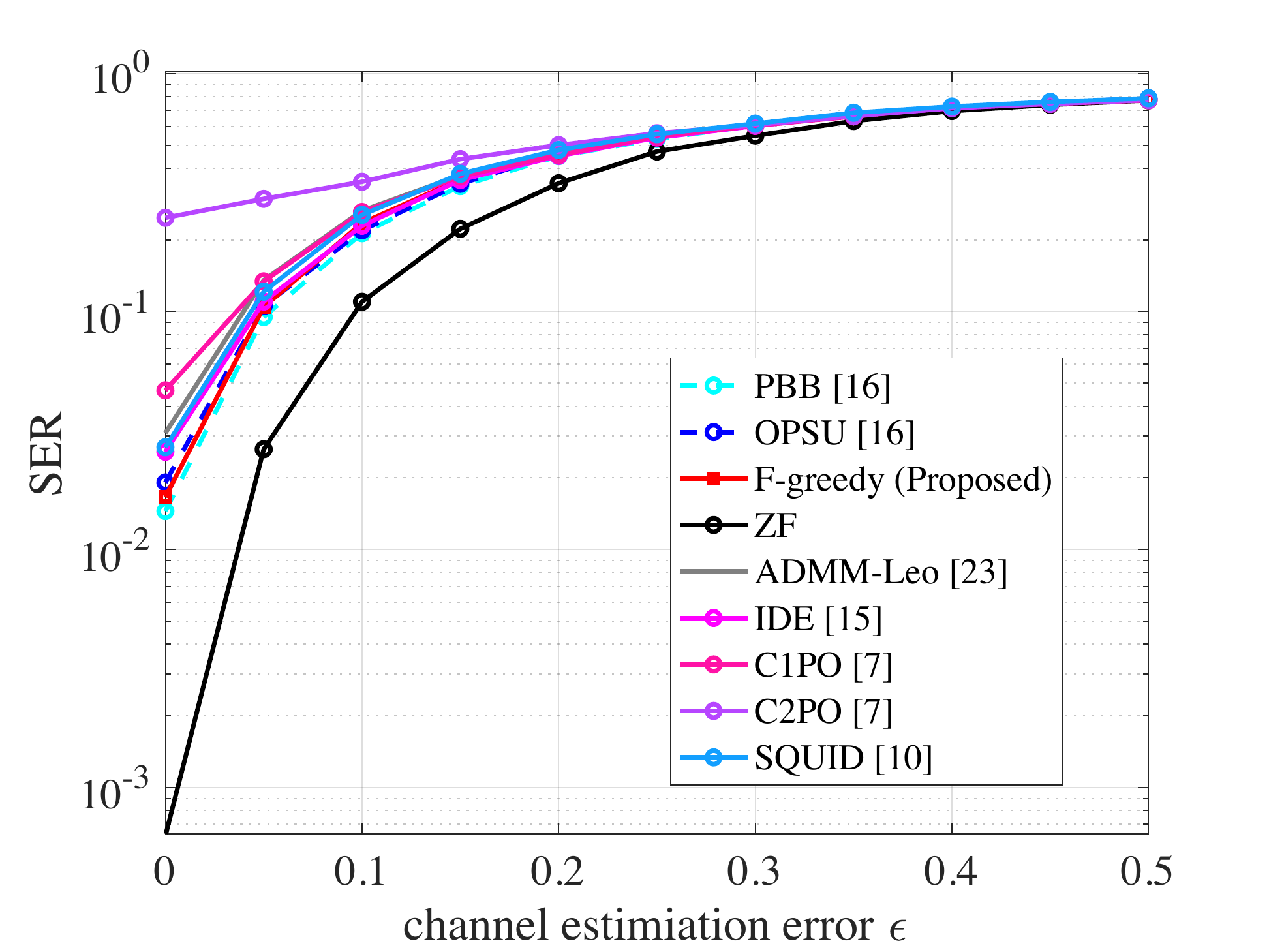}
     \caption{Performance comparisons of precoding methods for the downlink MU-MISO systems with 1-Bit DACs for channel estimiation error $\epsilon$, where $N_t$=64, $K$=8, and $4^2$-QAM with adaptive $\tau$ in $SNR$=10.}
     \label{fig:channel_error}
     \vspace{-1.0em}

\end{figure}

 We investigate the robustness of the proposed algorithm to channel estimation errors. We assume that the BS has access the imperfect CSI as
 \begin{equation}
     \Hm_{e} = \sqrt{1-\epsilon}\Hm + \sqrt{\epsilon}\Em,
 \end{equation}
 where $\epsilon\in[0,1]$ and $\Em\in\mathbbm{C}^{K\times N_t}$. Therefore, $\epsilon=1, \epsilon\in(0,1)$, and $\epsilon=0$ mean no CSI, partial CSI and perfect CSI scenarios, respectively. 
 In fig.~\ref{fig:channel_error}, the algorithm still achieve near-optimal performance with 10 {\rm dB} SNR under the imperfect CSI.  

\section{Conclusion}\label{sec:conclusion}
We have presented the construction of 1-bit transmit signal vector for downlink MU-MISO systems with QAM constellations. 
We define the linear feasibility conditions which guarantee that each user's noiseless received signal can be successfully detected as the desired message.
Also, the problem is transformed into the cascaded matrix form and further constructed as MILP problem. Solving MILP, 1-bit transmit signal vector with satisfying the feasibility conditions and the robustness to an additive noise. 
To efficiently solve MILP, we proposed the LP-relaxed algorithm that solve relaxed LP and refine the LP solution to satisfy 1-bit constraint.
Via numerical results, the proposed method is demonstrated superior performances with low-complexity compared with the benchmarks. 
For a more thorough discussion, please see \cite{park2021construction}.

\section*{Acknowledgment}
This work was supported by the National Research Foundation of Korea (NRF) grant funded by the Korea government (MSIT) (NRF-2020R1A2C1099836).



\bibliographystyle{IEEEtran}
\bibliography{journal_1bit_v3.bib}

\end{document}